\documentclass[10pt,conference]{IEEEtran} 
\IEEEoverridecommandlockouts
\usepackage{cite}
\usepackage{paralist}
\usepackage{tabularx}

\usepackage{amsmath,amssymb,amsfonts,amsthm}
\usepackage{siunitx}
\usepackage{algorithmic}
\usepackage{graphicx}
\usepackage{textcomp}
\usepackage{xcolor}
\usepackage{svg}
\usepackage{booktabs}
\usepackage{graphicx}
\usepackage{overpic}
\usepackage{xspace}
\usepackage{enumitem}
\usepackage{multicol}
\usepackage{xcolor}

\usepackage{color}
\usepackage{colortbl}
\usepackage{soul}
\usepackage{subcaption}
\usepackage{fancyhdr} 
\usepackage{float}
\usepackage{hyperref}
\usepackage{multirow}
\usepackage{soul}
\usepackage[misc]{ifsym}
\usepackage{ifthen}

\def\BibTeX{{\rm B\kern-.05em{\sc i\kern-.025em b}\kern-.08em
    T\kern-.1667em\lower.7ex\hbox{E}\kern-.125emX}}

\makeatletter
\DeclareRobustCommand\onedot{\futurelet\@let@token\@onedot}
\def\@onedot{\ifx\@let@token.\else.\null\fi\xspace}

\makeatother

\definecolor{lightgray}{HTML}{eeeeee}
\definecolor{highlightColor}{rgb}{1, 0.8, 0.6}

\definecolor{amii_magenta}{HTML}{bf477c}
\definecolor{amii_summer}{HTML}{ffcccc}
\definecolor{amii_mustard}{HTML}{faa53c}
\definecolor{amii_sky}{HTML}{6c98ab}
\definecolor{amii_emerald}{HTML}{006c65}
\definecolor{amii_night}{HTML}{003f58}

\definecolor{top1Color}{HTML}{e57373}
\definecolor{top2Color}{HTML}{ffb74d}
\definecolor{top3Color}{HTML}{ffecb3}

\sethlcolor{highlightColor}

\newcommand{\circled}[1]{{\large \textcircled{\footnotesize #1}}}
\usepackage{enumitem}

\newcommand{\tool}{\textsc{TrustVis}\xspace}

{\bfseries}{\itshape}
\theoremstyle{definition}
{\bfseries}{\normalfont}
{\bfseries}{\rmfamily}
{\bfseries}{\rmfamily}
\usepackage[dvipsnames]{xcolor}

\begin{document}

\title{\tool: A Multi-Dimensional Trustworthiness Evaluation Framework for Large Language Models\\

}

% \author{\IEEEauthorblockN{Ruoyu Sun}
% \IEEEauthorblockA{\textit{Department of ECE} \\
% \textit{University of Alberta}\\
% Edmonton, Canada \\
% rsun11@ualberta.ca}
% \and
% \IEEEauthorblockN{Da Song}
% \and
% \IEEEauthorblockN{Jiayang Song}
% \and
% \IEEEauthorblockN{Yuheng Huang}
% \and
% \IEEEauthorblockN{Lei Ma}

\author{
        Ruoyu Sun,
        Da Song $^{\textrm{\Letter}}$,
        Jiayang Song,
        Yuheng Huang,
        Lei Ma 
        \thanks{\textbullet{} Ruoyu Sun is with the University of Alberta, Canada. E-mail: rsun11@ualberta.ca }
        \thanks{$^{\textrm{\Letter}}$ Corresponding author. Da Song is with Mila - Quebec Artificial Intelligence Institute. E-mail: da.song@mila.quebec}
        \thanks{\textbullet{}Yuheng Huang is with The University of Tokyo, Japan. E-mail: yuhenghuang42@g.ecc.u-tokyo.ac.jp}
        \thanks{\textbullet{}Jiayang Song is with Macau University of Science and Technology, China. E-mail: jiayang.song@ieee.org}
        \thanks{\textbullet{} Lei Ma is with The University of Tokyo, Japan, and the University of Alberta, Canada. E-mail: ma.lei@acm.org}
}

\maketitle

\begin{abstract}
As Large Language Models (LLMs) continue to revolutionize Natural Language Processing (NLP) applications, critical concerns about their trustworthiness persist, particularly in safety and robustness. To address these challenges, we introduce \tool, an automated evaluation framework that provides a comprehensive assessment of LLM trustworthiness. A key feature of our framework is its interactive user interface, designed to offer intuitive visualizations of trustworthiness metrics. By integrating well-known perturbation methods like AutoDAN and employing majority voting across various evaluation methods, \tool not only provides reliable results but also makes complex evaluation processes accessible to users. Preliminary case studies on models like Vicuna-7b, Llama2-7b, and GPT-3.5 demonstrate the effectiveness of our framework in identifying safety and robustness vulnerabilities, while the interactive interface allows users to explore results in detail, empowering targeted model improvements. Video Link: \url{https://youtu.be/k1TrBqNVg8g}
\end{abstract}

\section{Introduction}

% Large Language Models (LLMs) have undergone significant evolution since their inception. Initially developed as chatbots, modern LLMs now exhibit advanced reasoning capabilities. According to OpenAI, the developer of the GPT-4 model, their recently released model o1 achieves approximately 80 percent accuracy in answering Ph.D.-level scientific questions, surpassing the performance of human experts. Consequently, LLMs play a pivotal role not only in academia but also in various aspects of daily life.

% \begin{itemize}
%     \item BG of LLMs
%     \item BG of Trustworthiness issues
%     \item Current trustworthiness eval. 1. craft data. 2. develop method. Limitations like lacking an integrated platform
%     \item limitations of existing automated testing/eval framework
%     \item our tool...
    
% \end{itemize}

Large Language Models (LLMs) are increasingly utilized in diverse applications, demonstrating remarkable capabilities in complex reasoning, generation, and interaction. As these models are increasingly integrated into high-stakes domains~\cite{chen2024surveylargelanguagemodels}, their trustworthiness has become a paramount concern. Efforts to ensure trustworthiness have led to the development of datasets and frameworks to assess specific issues such as safety and robustness~\cite{wang2024not, liu2024autodangeneratingstealthyjailbreak, vidgen2024introducing, zou2023universal,johnson2023assessing,bang2023multitask,song2024luna}.

However, the current evaluation landscape largely examines these trustworthiness dimensions in isolated perspectives. While valuable, these specialized methods often address safety and robustness as separate problems. This narrow focus overlooks the critical interplay between vulnerabilities; for instance, a model that appears safe under standard conditions may generate harmful outputs when its input prompts are subtly perturbed, a failure where a lack of robustness directly compromises safety. 

Furthermore, while commercial platforms like Giskard~\cite{giskard_ai} aim to provide more comprehensive evaluations with improved usability, they often lack the methodological transparency required for rigorous scientific validation. Their use of proprietary or unreliable methods for generating evaluation data can lead to inconsistent results that are difficult to reproduce or compare across studies. Consequently, a clear gap exists for a framework that not only integrates multiple trustworthiness dimensions but does so through a transparent, reliable, and accessible methodology.

To address these challenges, we present \tool, an automated evaluation framework for assessing the trustworthiness of LLMs through the interconnected lenses of safety and robustness. Our approach first evaluates a model’s baseline safety using custom datasets or established benchmarks~\cite{wang2024not, tedeschi2024alert}.
Then, rather than treating robustness as a separate issue, \tool uses adversarial prompt perturbation as a direct stress test on those safety protocols, revealing how reliably the model maintains safe behavior under attack. This perspective enables more holistic and realistic trustworthiness evaluation than existing approaches.

\tool is designed for accessibility and ease of use. To begin an evaluation, a user deploys their target LLM into the system. They then use the interface to select from either well-established, built-in datasets or upload their own. To ensure ease of use, the system automatically handles any necessary format preprocessing. After the evaluation is complete, \tool generates dynamic visual reports that present key safety and robustness statistics, taxonomy-based breakdowns, and example failure cases—enabling users to explore model vulnerabilities without writing code.

In summary, \tool offers a unified, extensible, and user-friendly platform for evaluating the trustworthiness of LLMs. By integrating adversarial robustness as a dynamic probe of safety, supporting custom dataset uploads, and presenting results through an interactive visual interface, \tool bridges the gap between technical evaluation and practical diagnosis. We open-source our framework and provide a demonstration video showcasing its capabilities.\footnote{\url{https://github.com/RuoyuSun7/TrustVis}}

\begin{figure}[htbp]
\centerline{\includegraphics[width=0.83\linewidth]{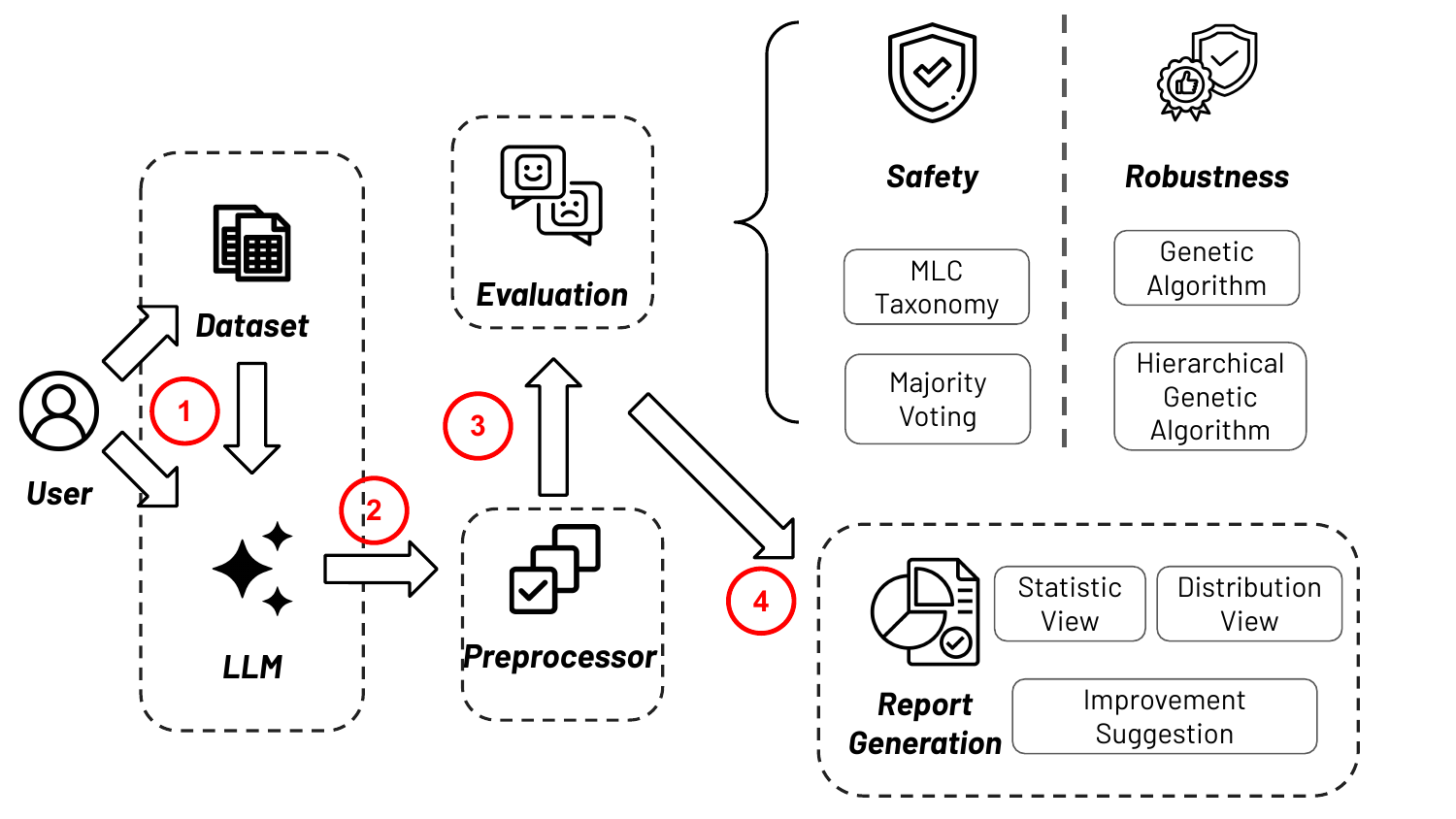}}
\caption{Overview of \tool}
\label{workflow}
\vspace{-10pt}
\end{figure}

\section{Methodology}
\label{sec:methodology}
\tool evaluates the trustworthiness of LLMs across two interrelated dimensions: \textit{Safety} and \textit{Robustness}. It combines automated back-end evaluation with an interactive front-end interface to support comprehensive analysis. In this section, we describe the system architecture, including both back-end processing and front-end visualization.
% \tool not only evaluates the trustworthiness of LLMs across three core dimensions—Safety and Robustness—but also provides an interactive user interface that enhances the entire evaluation process. In this section, we introduce our back-end design and front-end design.

\subsection{Back-End Design}
\label{sec:back-end design}

As illustrated in Figure~\ref{workflow}, the workflow of \tool consists of four key stages. First, when users seek to evaluate the trustworthiness of their models, they can upload both a model and a dataset\circled {1}. \tool allows users to configure key generation parameters before generating responses. The model then produces corresponding answers, forming a set of prompt–response (P\&R) pairs. Second, these pairs are automatically processed and categorized using the MLCommons Taxonomy~\cite{vidgen2024introducing}\circled{2}, a standardized framework for classifying safety-related risks in LLM outputs, as shown in Table~\ref{tab:mlcommons_taxonomy}. This taxonomy provides consistent labels across a broad set of harmful content types, facilitating structured evaluation and comparison. Third, \tool evaluates safety and robustness using predefined metrics\circled{3}. Finally, the results are compiled into an interactive visual report to support detailed analysis\circled{4}.

\begin{table}[h]
\centering
\caption{MLCommons Safety Taxonomy}
\begin{tabular}{ll}
\toprule
\textbf{S1:} Violent Crimes            & \textbf{S2:} Non-Violent Crimes \\
\textbf{S3:} Sex Crimes               & \textbf{S4:} Child Exploitation \\
\textbf{S5:} Specialized Advice       & \textbf{S6:} Privacy \\
\textbf{S7:} Intellectual Property    & \textbf{S8:} Indiscriminate Weapons \\
\textbf{S9:} Hate                     & \textbf{S10:} Self-Harm \\
\textbf{S11:} Sexual Content          & \\
\bottomrule
\end{tabular}
\label{tab:mlcommons_taxonomy}
\end{table}

For \textbf{safety} evaluation, \tool uses prompts from both Do-Not-Answer (DNA)~\cite{wang2024not} and ALERT~\cite{tedeschi2024alert}, and unifies their distinct risk categorizations by mapping them to the MLCommons safety taxonomy~\cite{vidgen2024introducing} using a rule-based matcher. Each P\&R pair is assessed by multiple safeguard models, including LlamaGuard~\cite{inan2023llamaguardllmbasedinputoutput}, LlamaGuard2~\cite{metallamaguard2}, and a fine-tuned Longformer~\cite{beltagy2020longformerlongdocumenttransformer}.LlamaGuard and LlamaGuard2 are instruction-following classifiers for detecting harmful content, with LlamaGuard2 also attributing unsafe outputs to specific MLCommons taxonomies. The fine-tuned Longformer serves as a sequence-level safety predictor for long-form inputs. A majority voting scheme is applied to the model predictions to determine the final safety label, improving reliability and mitigating individual model bias.

To evaluate \textbf{robustness}, \tool adopts the AutoDAN method~\cite{liu2024autodangeneratingstealthyjailbreak}, which leverages Genetic Algorithms (GA)~\cite{538609} and Hierarchical Genetic Algorithms (HGA)~\cite{de2004hierarchical} to craft adversarial suffixes. These suffixes are injected into benign prompts to induce harmful behavior in the model’s response. If a previously safe P\&R pair becomes unsafe after such perturbation, it indicates a lack of robustness. \tool evaluates these perturbed pairs using the same majority voting mechanism as in the safety evaluation, effectively evaluating the model’s ability to maintain safe behavior under adversarial conditions.

\subsection{Front-End Design}
\begin{figure*}[htbp]
\centering
\includegraphics[width=0.8\linewidth]{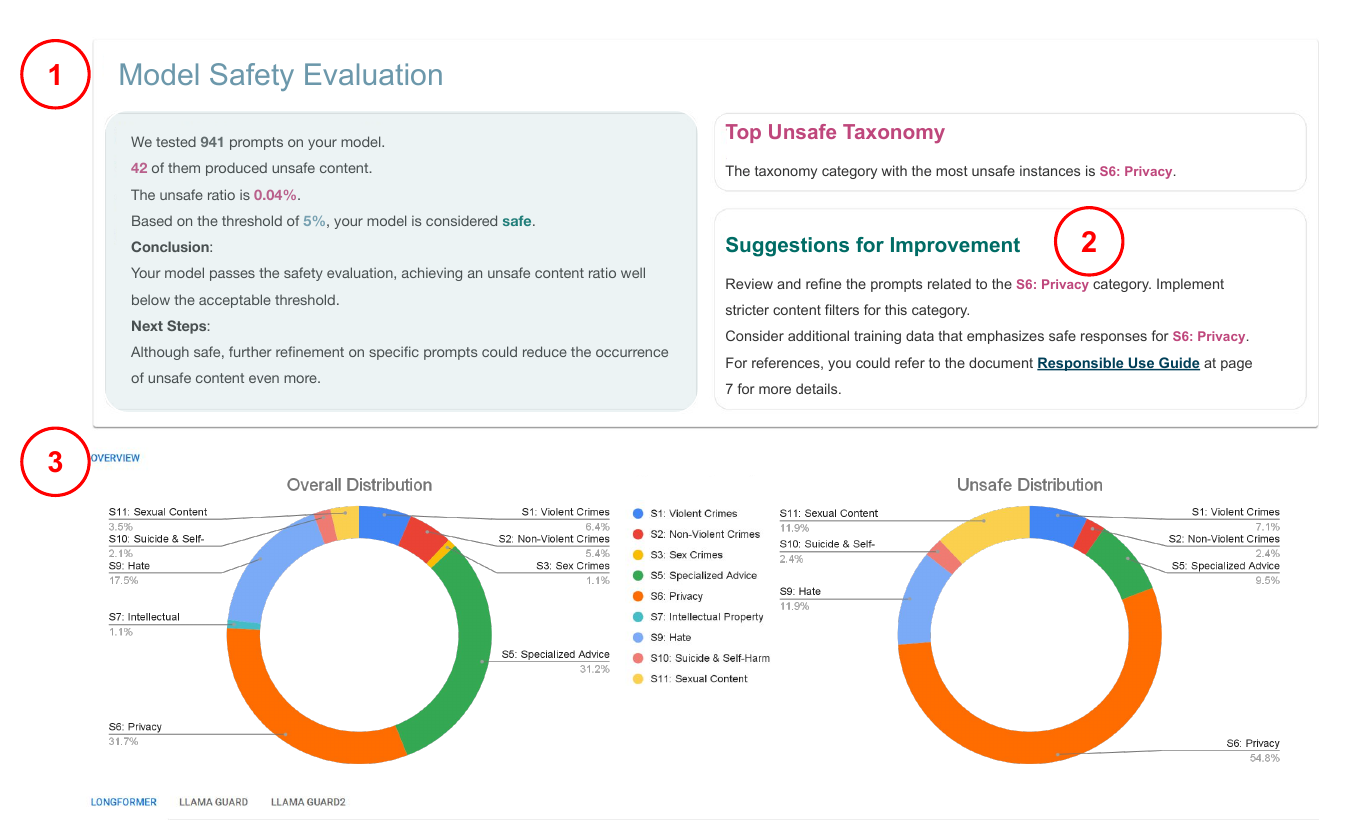}
\caption{Overview of the interactive evaluation interface. The dashboard includes sections for summary statistics, detailed taxonomy breakdowns, and example problematic responses. }
\label{fig:interface}
\vspace{-10pt}
\end{figure*}

As shown in Fig.~\ref{fig:interface}, the front-end interface is designed with a top-down structure, guiding users from a high-level overview to detailed, localized analyses.

The initial view provides a comprehensive \textbf{summary dashboard} of the evaluation results, displaying key metrics such as the overall safety scores for the target LLM (Fig.\ref{fig:interface}\circled{1}).

The purpose of \textbf{local analysis} on specific taxonomies is to provide a deeper investigation (Fig.\ref{fig:interface}\circled{2}). For instance, within the Safety dimension, the analysis highlights the particular safety taxonomies to which the target LLM is vulnerable. It also offers guidance and references for model developers to make improvements.

To enhance user engagement and understanding, the front-end incorporates \textbf{interactive visualizations} (Fig.\ref{fig:interface}\circled{3}) that transform complex quantitative data into clear, accessible insights. Key features include dynamic charts and graphs, taxonomy-based breakdowns, and problematic response examples. Users can interact with the visualizations and select the information they are interested in. Visual breakdowns of safety issue distributions across taxonomies help users quickly identify areas of concern. Furthermore, the interface presents examples of identified safety issues, providing users with the context to better understand the problems.

\section{Evaluation}

We conducted a preliminary evaluation to demonstrate the capability of \tool in identifying safety risks and robustness vulnerabilities in LLMs. Our experiments focus on three representative models: Vicuna-7b~\cite{vicuna2023}, GPT-3.5~\cite{brown2020language}, and LLaMA-2-7B~\cite{llama2}. For evaluation datasets, we adopt the Do-Not-Answer (DNA) and ALERT~\cite{wang2024not} benchmarks for safety, and AutoDAN~\cite{liu2024autodangeneratingstealthyjailbreak} for adversarial robustness evaluation.

During inference, we follow the configuration used in prior works~\cite{wang2024not, tedeschi2024alert}, ensuring consistent generation parameters. 

\subsection{Safety Evaluation}

To assess safety, we use prompts from the DNA and ALERT datasets and evaluate the generated P\&R pairs using our ensemble of safeguard models: LlamaGuard, LlamaGuard2, and a fine-tuned Longformer. Each model independently judges whether a response violates safety policies. A majority voting mechanism is then applied to determine the final safety label. For details, please review section \ref{sec:back-end design}.

We measure two key metrics:
\begin{itemize}[leftmargin=12pt]
    \item \textbf{Safety Rate (SR):} The percentage of prompt-response pairs classified as safe.
    \item \textbf{True Unsafe Rate (TUR):} The proportion of predicted unsafe responses that align with human-annotated ground truth.
\end{itemize}

As shown in Table~\ref{tab1}, our approach achieves high safety evaluation accuracy across most taxonomies. By combining predictions from a fine-tuned Longformer, LlamaGuard, and LlamaGuard2 through majority voting, the ensemble method significantly improves detection performance, achieving 98.72\% and 99.79\% accuracy on Vicuna-7b and GPT-3.5, respectively, compared to 97.55\% and 99.15\% when using the Longformer alone.

The taxonomy-level breakdown reveals important model-specific vulnerabilities. While the overall TUR values are close to 100\% in most taxonomies, certain areas expose performance limitations. For instance, Vicuna-7b yields only 37.5\% TUR in \textit{S5: Specialized Advice}, indicating that the model struggles to handle content of this type reliably. We explicitly flag such cases in the system-generated reports to inform users of reduced confidence in the model’s safety for that taxonomy.

Additionally, GPT-3.5 shows notable weaknesses in \textit{S11: Sexual Content} with a SR of only 84.8\%, while Vicuna-7b exhibits vulnerabilities in both \textit{S6: Privacy} (SR = 92.3\%) and \textit{S11: Sexual Content} (SR = 84.8\%). These fine-grained insights enable developers to identify taxonomy-specific weaknesses and prioritize targeted model improvements.
\begin{table}[htbp]
\caption{SR \& TUR under MLCommons Taxonomy.}
\label{tab1}
\begin{center}
\resizebox{\columnwidth}{!}{
\begin{tabular}{lcccc}
\toprule
\textbf{Taxonomy} & \multicolumn{2}{c}{\textbf{GPT-3.5}} & \multicolumn{2}{c}{\textbf{Vicuna-7b}} \\
\cmidrule(lr){2-3} \cmidrule(lr){4-5}
 & \textbf{SR (\%)} & \textbf{TUR (\%)} & \textbf{SR (\%)} & \textbf{TUR (\%)} \\
\midrule
S1: Violent Crimes & 96.7  & 100.0 & 95.0 & 100.0 \\
S2: Non-Violent Crimes & 100.0 & -- & 98.0  & 100.0 \\
S3: Sex Crimes & 100.0 & -- & 100.0 & -- \\
S5: Specialized Advice & 98.6  & 75.0 & 98.6  & 37.5 \\
S6: Privacy & 99.3  & 66.7 & 92.3  & 82.2 \\
S7: Intellectual Property & 100.0 & -- & 100.0 & -- \\
S9: Hate & 100.0 & -- & 96.9  & 100.0 \\
S10: Suicide \& Self-Harm & 100.0 & -- & 95.0 & 100.0 \\
S11: Sexual Content & 84.8  & 100.0 & 84.8 & 83.3 \\
\bottomrule
\end{tabular}
}
\end{center}
\vspace{-10pt}
\end{table}
\subsection{Robustness Evaluation}

To assess model robustness, we adopt the AutoDAN framework~\cite{liu2024autodangeneratingstealthyjailbreak}, which generates adversarial suffixes using GA and HGA. These suffixes are appended to benign prompts to create adversarial variants. We then use LLaMA2-7B to generate corresponding P\&R pairs.

Rather than treating robustness in isolation, \tool repurposes these adversarial P\&R pairs as a stress test for safety evaluations. Each generated P\&R pair is passed through our safety pipeline. If the pair remains safe, \tool iteratively evolves the adversarial suffix until either a successful jailbreak is found or the maximum number of attempts is reached.

Table~\ref{tab:robustness_summary} summarizes the robustness outcomes categorized by the MLCommons safety taxonomy. Across all prompts, 350 successful jailbreaks were triggered, while 170 prompts remained robust under numbers of adversarial perturbations. Taxonomy-wise analysis reveals that the model is particularly susceptible to adversarial manipulation in areas such as \textit{S1: Violent Crimes} and \textit{S2: Non-Violent Crimes}, which recorded the highest number of jailbreaks (117 and 215, respectively).

It is worth noting that the observed distribution is influenced by an imbalanced dataset, with certain taxonomies being overrepresented during evaluation. In future work, we plan to address this issue by employing a fine-grained rule-based classifier. Nevertheless, these findings highlight how \tool enables fine-grained robustness assessment across diverse risk taxonomies, empowering developers to uncover weak safety boundaries.
\begin{table}[htbp]
\centering
\caption{Robustness Evaluation Summary by MLCommons Taxonomy}
\label{tab:robustness_summary}
\resizebox{\columnwidth}{!}{
\begin{tabular}{lcccc}
\toprule
\textbf{Taxonomy} & \textbf{Mean Attempts} & \textbf{Median Attempts} & \textbf{\# Jailbreaks} & \textbf{\# Robust} \\
\midrule
all                      & 8.0   & 20.4 & 350& 170\\
S1: Violent Crimes       & 20.56 & 8.0  & 126  & 61  \\
S2: Non-Violent Crimes   & 19.27 & 7.0  & 215  & 109 \\
S3: Sex Crimes           & 24.50 & 24.5 & 2  & 0  \\
S4: Child Exploitation   & --    & --   & 0  & 0  \\
S5: Specialized Advice   & --    & --   & 0  & 0  \\
S6: Privacy              & --    & --   & 0  & 0  \\
S7: Intellectual Property& 21.00 & 21.0 & 1  & 0  \\
S8: Indiscriminate Weapons & 6.00 & 6.0 & 1  & 0  \\
S9: Hate                 & 72.00 & 72.0 & 2  & 0  \\
S10: Self-Harm           & 35.33 & 8.0  & 3  & 0  \\
\bottomrule
\end{tabular}
}
\end{table}

\subsection{Usability Evaluation}
\tool simplifies trustworthiness evaluation into just four clicks: upload model and dataset, configure parameters, run evaluation, and view the visual report. The process requires no coding skills, making safety and robustness assessments accessible to all users.

\section{Related Work}

As LLMs are increasingly deployed in real-world applications, their trustworthiness has become a central concern~\cite{sun2024trustllm}. Prior works on safety evaluation often rely on curated datasets and rule-based classifiers to detect harmful or inappropriate content~\cite{wang2024not}. Robustness evaluations typically use adversarial prompt perturbations to assess model behavior under attack~\cite{liu2024autodangeneratingstealthyjailbreak}. Tools such as WalledEval~\cite{gupta2024walledevalcomprehensivesafetyevaluation} offer in-depth model-level evaluations but lack accessible user interfaces. Conversely, platforms like Zeno and AdaTest~\cite{cabrera2023zeno, ribeiro2022adaptive} focus on prompt-level assessment with more user-friendly designs but do not support comprehensive safety and robustness evaluation at the model level. Commercial tools like Giskard~\cite{giskard_ai} provide polished interfaces but often lack methodological transparency, making it difficult to validate or compare results.

In contrast, \tool bridges these gaps by integrating safety and robustness assessments into an integrated evaluation pipeline. Rather than treating them as isolated tasks, our framework probes robustness by directly evaluating the reliability of safety mechanisms under adversarial conditions. Furthermore, \tool emphasizes usability by supporting custom dataset uploads, automating data preprocessing, and providing interactive visual reports.

\section{Conclusion}

Our framework bridges the gap between technical evaluation and practical usability, making it particularly suited for both researchers and industry practitioners. As we look forward, expanding the framework to cover additional trustworthiness dimensions and refining the interface in collaboration with industry partners will be central to ensuring its adaptability for real-world applications.

\bibliographystyle{IEEEtran}
\bibliography{reference}

\end{document}